\documentclass[seceq]{ptptex}
\title{%        %You can use \\ for explicit line-break
Limitation on Magnitude of $D$-components%
}

\author{%       %Use \scshape  for the family name
Yoshiharu \textsc{Kawamura}\footnote{E-mail:
haru@azusa.shinshu-u.ac.jp}%
}

\inst{%     %Affiliation, neglected when [addenda] or [errata]
Department of Physics, Shinshu University, Matsumoto 390-8621, Japan
}

\recdate{%      %Editorial Office will fill in this.
September 1, 2010}
\abst{%         %this abstract is neglected when [addenda] or [errata]
We study the magnitude of $D$-components in a generic supersymmetric field theory.
There exists $F$-component whose vacuum expectation value is comparable to
or bigger than that of $D$-component, in the absence of Fayet-Iliopoulos term,
the large hierarchy in the charge spectrum and strongly interacting higher-dimensional
couplings in the K\"ahler potential, %and the gauge kinetic function,
if contributions from other terms than $F$ and $D$-terms are negligible.
}

\begin{document}

\maketitle

\section{Introduction}

Much effort has been devoted to construct a realistic model
beyond the standard model (SM) based on supersymmetry (SUSY)
which is broken softly in our visible world.
The SUSY is broken by non-vanishing vacuum expectation values 
(VEVs) of some auxiliary fields ($F$ and/or $D$) in a SUSY breaking sector.
The breakdown of SUSY is mediated to our visible world by some messengers.
Then soft SUSY breaking parameters depend on the VEVs of $F$ and $D$,
reflecting on how to break SUSY and how to mediate the breakdown of SUSY.

Recently, the role of $D$-terms in the breakdown of SUSY has been attracted attention for general gauge mediation.\cite{IS,DKS}
The $D$-terms have also played an important role  
through the $D$-term contribution to scalar masses,\cite{D,HK} in various models,
e.g., SUSY grand unified theories,\cite{KMY,KMY2}
effective theories from string models\cite{KK,DPS,DGPS,K,HKKN,DV},
effects due to the kinetic mixing\cite{DKM},
the gauge mediation,\cite{AH,NTWY}
the anomaly mediation\cite{PR,JJ},
the mirage mediation\cite{CJ} and models with Dirac gauginos.\cite{FNW,NRSU}
Hence it would be useful to set a course of a model-building if we obtain constraints on 
the VEVs of $F$ and $D$ model-independently.

There is the theorem that if the VEV of all $F$-components vanish, i.e., 
$\langle F_I \rangle = \langle \partial W/\partial \phi^I \rangle = 0$
where $W$ is the superpotential and $\phi^I$ are scalar fields,
there exists a SUSY preserving solution satisfying the $D$-flat conditions, 
$\langle D^{\alpha} \rangle = \langle \phi^{\dagger}_I (T^{\alpha}\phi)^I \rangle = 0$.\cite{BDFS,WB}
It is known that the VEV of dominant $F$-component is comparable to or bigger than that of any $D$-components 
in most SUSY breaking solutions through the analysis of explicit models.
There are models that the VEV of dominant $D$-component can be bigger than that of any $F$-components
in the presence of Fayet-Iliopoulos (FI) term\cite{FI} or the large hierarchy in the charge spectrum.\cite{GRS}
It is interesting to know whether or not these features hold in a more generic framework of SUSY field theory.
This is the motivation of our work.

In this paper, we study the magnitude of $D$-components model-independently, that is,
without specifying the form of K\"ahler potential (matter kinetic function), superpotential
and gauge kinetic function.
In the next section, we consider a generic global SUSY field theory in the absence of FI term.
In \S 3, we extend our discussion to the case with FI term, soft SUSY breaking terms and the local SUSY in order.
In \S 4, we present conclusions and a discussion.

\section{Magnitude of $D$-component in global SUSY field theory}

Let us consider the global SUSY Lagrangian density,
\begin{eqnarray}
&~& \mathcal{L}_{\rm SUSY} = \int d^2\theta d^2\overline{\theta} K(\Phi^I, \Phi^{\dagger}_J, V)
+ \left[ \int d^2\theta W(\Phi^I) + h.c. \right]
\nonumber \\
&~& ~~~~~~~~~~~~ 
+ \left[ \frac{1}{4} \int d^2\theta f_{\alpha\beta}(\Phi^I) W^{\alpha} W^{\beta} + h.c. \right]~,
\label{L-SUSY}
\end{eqnarray}
where $\Phi^I$, $\Phi^{\dagger}_J$ and $V=V^{\alpha} T^{\alpha}$ are chiral scalar superfields, anti-chiral scalar superfields and vector superfields,
$T^\alpha$ are gauge transformation generators,
$h.c.$ stands for the hermitian conjugate and
$W^{\alpha}$ are chiral spinor superfields constructed from $V^{\alpha}$.
$K(\Phi^I, \Phi^{\dagger}_J, V)$, $W(\Phi^I)$ and $f_{\alpha\beta}(\Phi^I)$ are 
K\"ahler potential (matter kinetic function), superpotential and gauge kinetic function.
Both $K(\Phi^I, \Phi^{\dagger}_J, V)$ and $W(\Phi^I)$ are gauge invariant.
The last terms in the right hand side of (\ref{L-SUSY}) come from the following terms,
\begin{eqnarray}
\left[\frac{1}{2} \int d^2\theta \mbox{tr} \left(f(\Phi^I) (W^{\alpha}T^{\alpha}) (W^{\beta}T^{\beta})\right) + h.c.\right]~,
\label{L-SUSY3rd}
\end{eqnarray}
where tr represents the trace over the gauge generators.

The scalar potential is given by
\begin{eqnarray}
&~& V_{\rm SUSY} = - F^I K_I^J F_J - F^I \frac{\partial W}{\partial \phi^I} - F_J \frac{\partial \overline{W}}{\partial \phi^{\dagger}_J} 
  - \frac{1}{2} \mbox{Re} f_{\alpha \beta} D^{\alpha} D^{\beta}  - D^{\alpha} (K_I(T^{\alpha}\phi)^I)
\nonumber \\
&~& ~~~~~~~~ = \frac{\partial \overline{W}}{\partial \phi^{\dagger}_J} \left(K^{-1}\right)_J^I \frac{\partial W}{\partial \phi^I} 
+ \frac{1}{2} \left(\mbox{Re} f^{-1}\right)_{\alpha \beta} (K_I(T^{\alpha}\phi)^I)(K_J(T^{\beta}\phi)^J)~,
\label{V-SUSY}
\end{eqnarray}
where $F^I$, $F_J$ and $D^{\alpha}$ are auxiliary components in $\Phi^I$, $\Phi^{\dagger}_J$ and $V^{\alpha}$.
Here $K=K(\phi^I, \phi^{\dagger}_J)$, $W=W(\phi^I)$, $\overline{W} = \overline{W}(\phi^{\dagger}_J)$, $f_{\alpha\beta}=f_{\alpha\beta}(\phi^I)$,
$K_I=\partial K/\partial \phi^I$, 
$K_I^J=\partial^2 K/\partial \phi^I \partial \phi^{\dagger}_J$ etc.
The $\phi^I$ and $\phi^{\dagger}_J$ are scalar components in $\Phi^I$ and $\Phi^{\dagger}_J$, respectively.
$(\mbox{Re} f^{-1})_{\alpha \beta}$ and $(K^{-1})^I_J$ are the inverse matrices of $\mbox{Re} f_{\alpha \beta}$ and  
$K_I^J$, respectively.
The last equality in (\ref{V-SUSY}) is derived using the equations of motion,
\begin{eqnarray}
&~& F^I K_I^J + \frac{\partial \overline{W}}{\partial \phi^{\dagger}_J} = 0~,~~ 
K_I^J F_J + \frac{\partial W}{\partial \phi^I} = 0~,
\label{F}\\
&~& \mbox{Re} f_{\alpha \beta} D^{\beta}  + K_I(T^{\alpha}\phi)^I = 0~.
\label{D}
\end{eqnarray}
The scalar potential is rewritten down by
\begin{eqnarray}
V_{\rm SUSY} = F^I K_I^J F_J + \frac{1}{2} \mbox{Re} f_{\alpha \beta} D^{\alpha} D^{\beta}~,
\label{V-SUSY2}
\end{eqnarray}
where $F^I = - (K^{-1})_J^I \partial \overline{W}/\partial \phi^{\dagger}_J$, $F_J = - (K^{-1})_J^I \partial W/\partial \phi^I$
and $D^{\alpha} = - (\mbox{Re} f^{-1})_{\alpha \beta}$ $(K_I(T^{\beta}\phi)^I)$.

The derivative of $V_{\rm SUSY}$ by $\phi^{I'}$ is given by
\begin{eqnarray}
&~& \frac{\partial V_{\rm SUSY}}{\partial \phi^{I'}} 
= - F^I K_{II'}^J F_J - F^I \frac{\partial^2 W}{\partial \phi^I \partial \phi^{I'}}
\nonumber \\
&~& ~~~~~~~~~~~~~~  - \frac{1}{2} (\mbox{Re} f_{\alpha \beta})_{I'} D^{\alpha} D^{\beta} - (\phi^{\dagger} T^{\alpha})_I K^I_{I'} D^{\alpha}~,
\label{VI'}
\end{eqnarray}
using the identity derived from the gauge invariance of K\"ahler potential,
\begin{eqnarray}
K_I(T^{\alpha}\phi)^I = (\phi^{\dagger} T^{\alpha})_I K^I~.
\label{K-inv}
\end{eqnarray}
{}From the stationary condition $\langle \partial V_{\rm SUSY}/\partial \phi^{I'} \rangle = 0$,
we derive the formula:
\begin{eqnarray}
&~& \langle F^I \rangle \langle K_{II'}^J \rangle \langle F_J \rangle + \mu_{II'} \langle F^I \rangle
\nonumber \\
&~& ~~~~~~~~~~~ + \frac{1}{2} \langle (\mbox{Re} f_{\alpha \beta})_{I'} \rangle \langle D^{\alpha} \rangle \langle D^{\beta} \rangle 
 + \langle (\phi^{\dagger} T^{\alpha})_I \rangle \langle K^I_{I'} \rangle \langle D^{\alpha} \rangle = 0~,
\label{<VI'>}
\end{eqnarray}
where $\mu_{II'} \equiv \langle \partial^2 W/\partial \phi^I \partial \phi^{I'} \rangle$
is the SUSY mass coming from the superpotential.

By multiplying $(T^{\alpha'} \phi)^{I'}$ to (\ref{VI'}),
we obtain
\begin{eqnarray}
&~& \frac{\partial V_{\rm SUSY}}{\partial \phi^{I'}} (T^{\alpha'} \phi)^{I'} 
= - F^I (K_{I'}(T^{\alpha'}\phi)^{I'})_I^J F_J 
\nonumber \\
&~& ~~~~~~~~~~~~
- \frac{1}{2} (\mbox{Re} f_{\alpha \beta})_{I'} (T^{\alpha'} \phi)^{I'} D^{\alpha} D^{\beta} 
- (\phi^{\dagger} T^{\alpha})_I K^I_{I'} (T^{\alpha'} \phi)^{I'} D^{\alpha}~,
\label{VI'D}
\end{eqnarray}
where we use (\ref{K-inv}) and the identities derived from the gauge
invariance of the superpotential,
\begin{eqnarray}
\frac{\partial W}{\partial \phi^{I'}} (T^{\alpha'}\phi)^{I'} = 0~,~~
\frac{\partial W}{\partial \phi^I \partial \phi^{I'}} (T^{\alpha'}\phi)^{I'} + \frac{\partial W}{\partial \phi^{I'}} (T^{\alpha'})^{I'}_{I} = 0~.
\label{W-inv}
\end{eqnarray}
Taking its VEV and using the stationary condition, 
we derive the formula:
\begin{eqnarray}
&~& \langle F^I \rangle \left\langle (K_{I'}(T^{\alpha'}\phi)^{I'})_I^J \right\rangle \langle F_J \rangle 
\nonumber \\
&~& ~~~~~~~~~~~~
+ \frac{1}{2} \langle (\mbox{Re} f_{\alpha \beta})_{I'} \rangle \langle (T^{\alpha'} \phi)^{I'} \rangle \langle D^{\alpha} \rangle \langle D^{\beta} \rangle  
+ (\hat{M}_V^2)^{\alpha\alpha'} \langle D^{\alpha} \rangle = 0~,
\label{<VI'D>}
\end{eqnarray}
where $(\hat{M}_{V}^{2})^{\alpha \alpha'}= \langle (\phi^{\dagger} T^{\alpha})_I K^I_{I'} (T^{\alpha'} \phi)^{I'} \rangle$
is the mass matrix of the gauge bosons up to the normalization due to the gauge coupling constants.  
The formula (\ref{<VI'D>}) is a counterpart of (B.13) in Ref.\citen{PS}.

By multiplying $(K^{-1})^{I'}_{I''} K^{I''}$ to (\ref{VI'}), 
we obtain
\begin{eqnarray}
&~& \frac{\partial V_{\rm SUSY}}{\partial \phi^{I'}} (K^{-1})^{I'}_{I''} K^{I''}
\nonumber \\
&~& ~~~~~~~ = - F^I K_{II'}^J F_J (K^{-1})^{I'}_{I''} K^{I''} - F^I \frac{\partial^2 W}{\partial \phi^I \partial \phi^{I'}} (K^{-1})^{I'}_{I''} K^{I''}
\nonumber \\
&~& ~~~~~~~~~~ - \frac{1}{2} (\mbox{Re} f_{\alpha \beta})_{I'}(K^{-1})^{I'}_{I''} K^{I''} D^{\alpha} D^{\beta} 
+ \mbox{Re} f_{\alpha \beta} D^{\alpha} D^{\beta}~.
\label{VI'K}
\end{eqnarray}
The relation (\ref{VI'K}) is a counterpart of the identity (4.5) in Ref.\citen{DKS}.
Taking its VEV and using the stationary condition, 
we derive the formula:
\begin{eqnarray}
&~& \langle F^I \rangle \langle K_{II'}^J \rangle \langle F_J \rangle \langle (K^{-1})^{I'}_{I''} \rangle \langle K^{I''} \rangle 
+ \mu_{II'} \langle F^I \rangle \langle (K^{-1})^{I'}_{I''} \rangle \langle K^{I''} \rangle
\nonumber \\
&~& ~~~ + \frac{1}{2} \langle (\mbox{Re} f_{\alpha \beta})_{I'} \rangle \langle (K^{-1})^{I'}_{I''} \rangle 
\langle K^{I''} \rangle \langle D^{\alpha} \rangle \langle D^{\beta} \rangle 
= \langle \mbox{Re} f_{\alpha \beta} \rangle \langle D^{\alpha} \rangle \langle D^{\beta} \rangle~.
\label{<VI'K>}
\end{eqnarray}

We rearrange fields into those forming irreducible representations such as $(T^{\alpha})_I^J = T^{\alpha}_{(I)} \delta_I^J$
under gauge groups 
where $T^{\alpha}_{(I)}$ is the representation matrix for $\Phi^I$ and the same notation for $\Phi^I$ is used.
In $K=K(\phi^I, \phi^{\dagger}_J)$, fields with a same representation can be mixed such that
\begin{eqnarray}
K = a_I^J \phi^{\dagger}_J \phi^I + \frac{a_{II'}^J}{\Lambda} \phi^{\dagger}_J \phi^{I} \phi^{I'} + \cdots~,
\label{K-exp}
\end{eqnarray}
where $a_I^J$ and $a_{II'}^J$ are coefficients and $\Lambda$ is a high energy scale.
The VEV  of $K_I^J$ is estimated as 
\begin{eqnarray}
&~& \langle K_I^J \rangle = a_I^J + \frac{a_{II'}^J}{\Lambda} \langle \phi^{I'} \rangle + \frac{a_{I'I}^J}{\Lambda} \langle \phi^{I'} \rangle + \cdots
\nonumber \\
&~& ~~~~~~~ = a_I^J + O(\langle \phi^{I'} \rangle/\Lambda)~,
\label{<KIJ>}
\end{eqnarray}
where we assume that the magnitude of $a_{II'}^J$ and higher coefficients is at most $O(1)$
and the magnitude of $\langle \phi^{I'} \rangle$ is comparable to or less than $\Lambda$. 

The non-vanishing VEV of $D$-component implies the breakdown of gauge symmetry
by the VEV of some gauge non-singlet scalar fields,
in the absence of  FI term.
The non-vanishing components in $\langle D \rangle \equiv \langle D^{\alpha} \rangle T^{\alpha}$ are those
for diagonal generators $T^a$ because $\langle D \rangle$ is transformed into $\langle D^a \rangle T^a$
by some unitary matrix $U$.
Because the fields forming a same representation change in a same manner under the unitary transformation,
the form of $K$ is invariant after the redefinition of fields by $U$ and we use the same notation for fields
to avoid confusion.
The VEV of $D^a$ is written by
\begin{eqnarray}
\hspace{-0.8cm}&~& \langle D^a \rangle = - \langle (\mbox{Re} f^{-1})_{aa} \rangle \langle K_I (T^{a} \phi)^I \rangle 
= - g_a^2 q^a_{\tiny{(\phi^I)}} \langle K_I \phi^I \rangle
\nonumber \\
\hspace{-0.8cm}&~& ~~~~~~ = - g_a^2 q^a_{\tiny{(\phi^I)}} \left(a_I^J \langle \phi^{\dagger}_J \rangle \langle \phi^I \rangle 
+ \frac{a_{II'}^J}{\Lambda} \langle \phi^{\dagger}_J \rangle \langle \phi^I \rangle \langle \phi^{I'} \rangle + \cdots\right)
\nonumber \\
\hspace{-0.8cm}&~& ~~~~~~ 
= - g_a^2 q^a_{\tiny{(\phi^I)}} \left(\!\!\left(\langle K_I^J \rangle + O(\langle \phi^{I'} \rangle/\Lambda)\right)\!\!
\langle \phi^{\dagger}_J \rangle \langle \phi^I \rangle 
+ \frac{a_{II'}^J}{\Lambda} \langle \phi^{\dagger}_J \rangle \langle \phi^I \rangle \langle \phi^{I'} \rangle + \cdots\right),
\label{<Da>}
\end{eqnarray}
where $q^a_{\tiny{(\phi^I)}}$ is the value of $T^a_{(I)}$ for the non-vanishing component of $\phi^I$
and the gauge coupling constant $g_a$ is given by $g_a^2 = \langle (\mbox{Re} f^{-1})_{aa} \rangle$.
We assume that the magnitude of $\langle K_I^J \rangle$ is $O(1)$.
After the diagonalization of $\langle K_I^J \rangle$, $\langle D^a \rangle$ is written by
\begin{eqnarray}
\langle D^a \rangle = - g_a^2 q^a_{\tiny{(\phi^I)}} \left|\langle \phi^I \rangle\right|^2 \left(1 + O(\langle \phi^{I'} \rangle/\Lambda)\right)~,
\label{<Da2>}
\end{eqnarray}
where we also use the same notation for fields after their redefinition.
Then the mass matrix of gauge bosons is diagonalized and the mass of gauge boson for $T^a$ is given by
\begin{eqnarray}
({M}_V^2)^a = g_a^2 (\hat{M}_V^2)^a = g_a^2 (q^a_{\tiny{(\phi^I)}})^2 \left|\langle \phi^I \rangle\right|^2~.
\label{MV2a}
\end{eqnarray}
%where we use the feature that $\langle K_I^J \rangle$ vanishes for fields with a different representation.

The first term in the left hand side of  (\ref{<VI'D>}) for the diagonal generator $T^a$ is written by
\begin{eqnarray}
&~& \langle F^I \rangle \left\langle (K_{I'}(T^a \phi)^{I'})_I^J \right\rangle \langle F_J \rangle 
\nonumber \\
&~& ~~~~ = \langle F^I \rangle \langle K_{I'}^J \rangle (T^a)_{I'}^J  \langle F_J \rangle 
+ \langle F^I \rangle \langle K_{I'I}^J \rangle \langle (T^a \phi)^{I'} \rangle \langle F_J \rangle
\nonumber \\
&~& ~~~~ = q^a_{\tiny{(F^I)}}  \left|\langle F^I \rangle\right|^2 
+ \langle F^I \rangle \langle K_{I'I}^J \rangle q^a_{\tiny{(\phi^{I'})}} \langle \phi^{I'} \rangle \langle F_J \rangle
\nonumber \\
&~& ~~~~ = q^a_{\tiny{(F^I)}}  \left|\langle F^I \rangle\right|^2 
\left(1 + O\left(\frac{q^a_{\tiny{(\phi^{I'})}}}{q^a_{\tiny{(F^I)}}} \frac{\langle \phi^{I'} \rangle}{\Lambda}\right)\right)~,
\label{1st<VI'D>}
\end{eqnarray}
where $q^a_{\tiny{(F^I)}}$ is the value of $T^a_{(I)}$ for the non-vanishing component of $F^I$.

The second term in the left hand side of  (\ref{<VI'D>}) vanishes for $T^a$
because the relation $\langle (\mbox{Re} f_{bc})_{I} \rangle \langle (T^{a} \phi)^{I} \rangle =0$
holds from the gauge invariance of $f_{bc}(\Phi^I)$.
Here $a$, $b$ and $c$ are indices for the Cartan sub-algebra.
Notice that $f_{bc}(\Phi^I)$, $D^b$ and $D^c$ are neutral under the $U(1)$ charges relating the Cartan sub-algebra.

Using (\ref{<Da2>}) and (\ref{1st<VI'D>}), 
the magnitude of $\langle D^a \rangle$ and $\langle F^I \rangle \left\langle (K_{I'}(T^a \phi)^{I'})_I^J \right\rangle \langle F_J \rangle$
are bounded as
\begin{eqnarray}
|\langle {D}^a \rangle| \le g_a^2 |q^a_{\tiny{(\phi^I)}}| \left|\langle \phi^I \rangle\right|^2
\left|1 + O(\langle \phi^{I'} \rangle/\Lambda)\right|
\label{mag<D>}
\end{eqnarray}
and 
\begin{eqnarray}
\langle F^I \rangle \left\langle (K_{I'}(T^a \phi)^{I'})_I^J \right\rangle \langle F_J \rangle
\le |q^a_{\tiny{(F^I)}}| \left|\langle F^I \rangle\right|^2 
\left|1 + O\left(\frac{q^a_{\tiny{(\phi^{I'})}}}{q^a_{\tiny{(F^I)}}} \frac{\langle \phi^{I'} \rangle}{\Lambda}\right)\right|~,
\label{mag<F>^2}
\end{eqnarray}
respectively.
Using (\ref{<VI'D>}), (\ref{MV2a}), (\ref{mag<D>}) and (\ref{mag<F>^2}), the magnitude of $\langle D^a \rangle^2$
is bounded as
\begin{eqnarray}
&~& \overline{q^a_{\tiny{(\phi)}}} \langle {D}^a \rangle^2 
\le ({M}_V^2)^a |\langle {D}^a \rangle|  \left|1 + O(\langle \phi^{I'} \rangle/\Lambda)\right|
\nonumber \\
&~& ~~~~~~~~~~~~ \le g_a^2 |q^a_{\tiny{(F^I)}}| \left|\langle F^I \rangle\right|^2 
\left|1 + O(\langle \phi^{I'} \rangle/\Lambda)
+ O\left(\frac{q^a_{\tiny{(\phi^{I'})}}}{q^a_{\tiny{(F^I)}}} \frac{\langle \phi^{I'} \rangle}{\Lambda}\right)\right|~,
\label{<D><<F>}
\end{eqnarray}
where $\overline{q^a_{\tiny{(\phi)}}}$ is defined by
\begin{eqnarray}
\overline{q^a_{\tiny{(\phi)}}} \equiv \frac{(\hat{M}_V^2)^a}{|q^a_{(I)}(\phi)| \left|\langle \phi^I \rangle\right|^2}
= \frac{(q^a_{\tiny{(\phi^I)}})^2 \left|\langle \phi^I \rangle\right|^2}{|q^a_{(I)}(\phi)| \left|\langle \phi^I \rangle\right|^2}~.
\label{|q|}
\end{eqnarray}
Eq.(\ref{<D><<F>}) is our master formula and,
from (\ref{<D><<F>}), we find that {\it the magnitude of $\langle {D}^{a} \rangle$ is comparable to}\footnote{
There are several models which generate comparable $\langle {D}^{a} \rangle$ and $\langle F^I \rangle$.\cite{DNNS,CFK,M,DKS}
}
{\it or smaller than that of dominant $\langle F^I \rangle$}\footnote{
The number of $\langle F^I \rangle$ contributing SUSY breaking dominantly is supposed to be not so large.
}
{\it if the condition $\overline{q^a_{\tiny{(\phi)}}} \ge O(g_a^2 |q^a_{\tiny{(F^I)}}|)$ is fulfilled.}
Here we restate our basic assumptions:
\begin{eqnarray}
%&~& g_{a}^2 \le O(1)~,~~ |q^a_{\tiny{(F^I)}}| = O\left(|q^a_{\tiny{(\phi^{I'})}}|\right)~,~~ 
\left|\langle \phi^I \rangle\right| \le \Lambda~,~~
\left|\langle K_{I_1 I_2 \cdots I_n}^{J_1 J_2 \cdots J_m} \rangle\right| = O\left(\frac{1}{\Lambda^{n+m-2}}\right)~~~ (n+m \ge 2)~.
%\langle (\mbox{Re} f_{\alpha \beta})_{I_1 I_2 \cdots I_n}^{J_1 J_2 \cdots J_m} \rangle = O\left(\frac{1}{\Lambda^{n+m-2}}\right)
\label{gqKf}
\end{eqnarray}
%Here and hereafter most values mean those absolute values.
%The first assumption states that the gauge symmetry is broken in a weak coupling regime if it occurs.
%The second one is that there is no large hierarchy between the broken charge of dominant SUSY breaking
%$F$-component and the broken charge of dominant gauge symmetry breaking scalar component.
These mean that the breakdown of gauge symmetry occurs below the scale $\Lambda$
and there are no strongly interacting higher-dimensional couplings in $K$, respectively.
%and $f_{\alpha \beta}$, respectively.

In the case with $g_a^2 |q^a_{\tiny{(F^I)}}| \gg \overline{q^a_{\tiny{(\phi)}}}$,  
the magnitude of $\langle {D}^a \rangle$ can be much bigger than that of $\langle F^I \rangle$
if the equalities in (\ref{<D><<F>}) hold approximately.
Actually an explicit model has been constructed with the large hierarchy in the charge spectrum.\cite{GRS}
We explain it briefly.
Let us take the O' Raifertaigh model with the following superpotential $W$,
\begin{eqnarray}
W = \lambda_1 \Phi_0 (\Phi_1 \Phi_{-1/N}^N -1) + \lambda_2 \Phi_1 \Phi_{-1} + \lambda_3 \Phi'_0 \Phi_{1/N} \Phi_{-1/N}~,
\label{exW}
\end{eqnarray}
where $\Phi_0$, $\Phi'_0$, $\Phi_1$, $\Phi_{-1}$ $\Phi_{1/N}$ and $\Phi_{-1/N}$ are 
chiral superfields with $U(1)$ charges $0$, $0$, $1$, $-1$, $1/N$ and $-1/N$.
The relation $|\langle D \rangle|^2 \sim N |\langle F^I \rangle|^2$ is derived,
and it leads to $|\langle D \rangle| \gg |\langle F^I \rangle|$ if $\sqrt{N} \gg 1$.
Here $D$ is the $D$-component of $U(1)$.
As the relation suggests, $\langle F_1 \rangle$ dominates in $\langle F^I \rangle$ and 
$\langle \phi_{-1/N} \rangle$ dominates in $\langle D \rangle$.

In the case with $|\langle K_{I'I}^J \rangle \langle \phi^{I'} \rangle| \gg O(1)$,
the term $\langle F^I \rangle \langle K_{I'I}^J \rangle \langle (T^a \phi)^{I'} \rangle \langle F_J \rangle$
dominates in $\langle F^I \rangle \left\langle (K_{I'}(T^a \phi)^{I'})_I^J \right\rangle \langle F_J \rangle$
and $\langle {D}^a \rangle^2$ is bounded as
\begin{eqnarray}
\overline{q^a_{\tiny{(\phi)}}} \langle {D}^a \rangle^2 
\le g_a^2 \left|\langle K_{I'I}^J \rangle q^a_{\tiny{(\phi^{I'})}} \langle \phi^{I'} \rangle \langle F^I \rangle \langle F_J \rangle\right|~.
\label{<D>><F>}
\end{eqnarray}
Then  the magnitude of $\langle {D}^a \rangle$ can be much bigger than that of $\langle F^I \rangle$
if the equality in (\ref{<D>><F>}) holds approximately 
and $g_a^2 |\langle K_{I'I}^J \rangle q^a_{\tiny{(\phi^{I'})}} \langle \phi^{I'} \rangle| \gg \overline{q^a_{\tiny{(\phi)}}}$.

In the case that all $F$-components vanish, we obtain the relation
\begin{eqnarray}
\frac{1}{2} \langle (\mbox{Re} f_{\alpha \beta})_{I'} \rangle \langle D^{\alpha} \rangle \langle D^{\beta} \rangle 
 + \langle (\phi^{\dagger} T^{\alpha})_I \rangle \langle K^I_{I'} \rangle \langle D^{\alpha} \rangle = 0~,
\label{<VI'>F=0}
\end{eqnarray}
from (\ref{<VI'>}) or 
\begin{eqnarray}
\frac{1}{2} \langle (\mbox{Re} f_{\alpha \beta})_{I'} \rangle \langle (K^{-1})^{I'}_{I''} \rangle 
\langle K^{I''} \rangle \langle D^{\alpha} \rangle \langle D^{\beta} \rangle 
= \langle \mbox{Re} f_{\alpha \beta} \rangle \langle D^{\alpha} \rangle \langle D^{\beta} \rangle~,
\label{<VI'K>F=0}
\end{eqnarray}
from (\ref{<VI'K>}).
Unless $\langle \mbox{Re} f_{\alpha \beta} \rangle$ equals to 
$\frac{1}{2} \langle (\mbox{Re} f_{\alpha \beta})_{I'} \rangle \langle (K^{-1})^{I'}_{I''} \rangle \langle K^{I''} \rangle$,\footnote{
As an example,
the relation $\langle \mbox{Re} f_{\alpha \beta} \rangle
= \frac{1}{2} \langle (\mbox{Re} f_{\alpha \beta})_{I'} \rangle \langle (K^{-1})^{I'}_{I''} \rangle \langle K^{I''} \rangle$
holds for the canonical K\"ahler potential $K = |\phi^I|^2$ and the non-minimal gauge kinetic function 
$f_{\alpha \beta} = c_{\alpha\beta} (\phi^I)^2$.
}
the $D$-flat conditions, $\langle D^{\alpha} \rangle = 0$, are derived
and then the SUSY is unbroken.

In this way, we obtain the following results.
\begin{itemize}
\item[(1)] The magnitude of $\langle {D}^{\alpha} \rangle$ is comparable to or smaller than that of dominant $\langle F^I \rangle$
under the assumptions (\ref{gqKf}),
unless the magnitude of the broken charge of $F$-components contributing SUSY breaking 
is much bigger than that of the broken charge of scalar components contributing gauge symmetry breaking.
\item[(2)] There always exists a SUSY vacuum in the case that all $F$-components vanish
and $\langle \mbox{Re} f_{\alpha \beta} \rangle$ is different from 
$\frac{1}{2} \langle (\mbox{Re} f_{\alpha \beta})_{I'} \rangle \langle (K^{-1})^{I'}_{I''} \rangle \langle K^{I''} \rangle$. 
\end{itemize}

\section{Several extensions}

We extend our discussion to several cases.

\subsection{Case with FI term}

For $U(1)$ gauge symmetries, the following term called Fayet-Iliopoulos term can be added to $\mathcal{L}_{\rm SUSY}$,
\begin{eqnarray}
\mathcal{L}_{\rm FI} = \int d^2\theta d^2\overline{\theta}  \xi_r V^r = \xi_r D^r~,
\label{L-FI}
\end{eqnarray}
where $\xi_r$ are constants, $V^r$ are $U(1)$ vector superfields and
$D^r$ are the auxiliary components in $V^r$.
The equations of motions for $D$-components are modified as
\begin{eqnarray}
\mbox{Re} f_{\alpha \beta} D^{\beta}  + K_I(T^{\alpha}\phi)^I + \xi_r \delta^{\alpha r} = 0~.
\label{DFI}
\end{eqnarray}
Then the scalar potential is modified as
\begin{eqnarray}
&~& V_{\rm SUSY} = - F^I K_I^J F_J - F^I \frac{\partial W}{\partial \phi^I} - F_J \frac{\partial \overline{W}}{\partial \phi^{\dagger}_J} 
\nonumber \\
&~& ~~~~~~~~~~~~~ 
 - \frac{1}{2} \mbox{Re} f_{\alpha \beta} D^{\alpha} D^{\beta}
 - D^{\alpha} \left(K_I(T^{\alpha}\phi)^I + \xi_r \delta^{\alpha r}\right)
\nonumber \\
&~& ~~~~~~~~ = \frac{\partial \overline{W}}{\partial \phi^{\dagger}_J} \left(K^{-1}\right)_J^I \frac{\partial W}{\partial \phi^I} 
\nonumber \\
&~& ~~~~~~~~~~~~~ 
+ \frac{1}{2} \left(\mbox{Re} f^{-1}\right)_{\alpha \beta} (K_I(T^{\alpha}\phi)^I + \xi_r \delta^{\alpha r})
 (K_I(T^{\beta}\phi)^I + \xi_r \delta^{\beta r})~.
\label{V-SUSYFI}
\end{eqnarray}
Although the same types of formulae (\ref{<VI'>}) and (\ref{<VI'D>}) are derived, 
the inequalities on $\langle D^r \rangle^2$ are different from (\ref{<D><<F>}) such that 
\begin{eqnarray}
&~& \overline{q^r_{\tiny{(\phi)}}} \langle {D}^r \rangle^2 
\le \eta_r ({M}_V^2)^r |\langle {D}^r \rangle|  \left|1 + O(\langle \phi^{I'} \rangle/\Lambda)\right|
\nonumber \\
&~& ~~~~~~~~~~~~ \le \eta_r g_r^2 |q^r_{\tiny{(F^I)}}| \left|\langle F^I \rangle\right|^2 
\left|1 + O(\langle \phi^{I'} \rangle/\Lambda)
+ O\left(\frac{q^r_{\tiny{(\phi^{I'})}}}{q^r_{\tiny{(F^I)}}} \frac{\langle \phi^{I'} \rangle}{\Lambda}\right)\right|~,
\label{<D><<F>FI}
\end{eqnarray}
where $g_r^2 = \langle (\mbox{Re} f^{-1})_{rr} \rangle$, and $\overline{q^r_{\tiny{(\phi)}}}$ and $\eta_r$ are defined by
\begin{eqnarray}
\overline{q^r_{\tiny{(\phi)}}} \equiv \frac{(\hat{M}_V^2)^r}{|q^r_{(I)}(\phi)| \left|\langle \phi^I \rangle\right|^2}
= \frac{(q^r_{\tiny{(\phi^I)}})^2 \left|\langle \phi^I \rangle\right|^2}{|q^r_{\tiny{(\phi^I)}}| \left|\langle \phi^I \rangle\right|^2}
\label{|qr|}
\end{eqnarray}
and 
\begin{eqnarray}
\hspace{-2cm} 
\eta_r \equiv \frac{|q^r_{\tiny{(\phi^I)}}| \left|\langle \phi^I \rangle\right|^2 + |\xi_r|}{|q^r_{\tiny{(\phi^I)}}| \left|\langle \phi^I \rangle\right|^2}~,
\label{etar}
\end{eqnarray}
respectively.
Here $q^r_{\tiny{(\phi^I)}}$ and $q^r_{\tiny{(F^I)}}$ are values of $T^r_{(I)}$ 
for the non-vanishing components of $\phi^I$ and $F^I$, respectively.
In the case with $\eta_r = O(1)$, the same result (1) is obtained.
If $\eta_r \gg 1$\footnote{
In an extremal case, there is a possibility that the VEV of $D^r$ is $\xi_r$ itself 
and non-vanishing but the $U(1)$ gauge symmetry is not broken with $\langle (T^{r}\phi)^I \rangle = 0$
and $\langle F^I \rangle = 0$, 
where $T^r$ is the $U(1)$ charge operator. 
} 
and the equalities in (\ref{<D><<F>FI}) hold approximately, 
the magnitude of $\langle D^r \rangle$ can be much bigger than that of $\langle F^I \rangle$
such that
\begin{eqnarray}
|\langle {D}^r \rangle| = O(|\xi_r|) \gg |\langle F^I \rangle|~.
\label{<D>><F>FI}
\end{eqnarray}

In the case that all $F$-components vanish, we obtain the relation (\ref{<VI'>F=0}) or 
\begin{eqnarray}
\frac{1}{2} \langle (\mbox{Re} f_{\alpha \beta})_{I'} \rangle \langle (K^{-1})^{I'}_{I''} \rangle 
\langle K^{I''} \rangle \langle D^{\alpha} \rangle \langle D^{\beta} \rangle 
+ \langle (\phi^{\dagger} T^{\alpha})_I K^I \rangle \langle D^{\alpha} \rangle = 0~.
\label{<VI'K>F=0FI}
\end{eqnarray}
There can appear a non-SUSY vacuum with $\langle D^r \rangle \ne 0$, in which the gauge symmetry is unbroken
with $\langle (\phi^{\dagger} T^r)_I \rangle = 0$, in the case that
$\langle (\mbox{Re} f_{rr'})_{I'} \rangle = 0$ and all $F$-components vanish
with $\langle (\phi^{\dagger} T^r)_I \rangle = 0$.

\subsection{Case with soft SUSY breaking terms}

In the case that SUSY is broken in other sector at some high-energy scale,
soft SUSY breaking terms can appear after mediating by some messengers.
We consider the following type of soft SUSY breaking terms for the scalar potential,\footnote{
The form of $U(\phi^I)$ is constrained by requiring that the gauge hierarchy achieved 
by a fine-tuning in the superpotential should not be violated by soft SUSY breaking terms.\cite{KMY2}
}
\begin{eqnarray}
{V}_{\rm soft} = (m^2)_I^J \phi^{\dagger}_J \phi^I + \left[U(\phi^I) + h.c.\right]~.
\label{V-soft}
\end{eqnarray}
In the presence of $V_{\rm soft}$, (\ref{<VI'>}) and (\ref{<VI'D>}) are modified as
\begin{eqnarray}
&~& \langle F^I \rangle \langle K_{II'}^J \rangle \langle F_J \rangle + \mu_{II'} \langle F^I \rangle
+ \frac{1}{2} \langle (\mbox{Re} f_{\alpha \beta})_{I'} \rangle \langle D^{\alpha} \rangle \langle D^{\beta} \rangle 
\nonumber \\
&~& ~~~~~~~~~~~  + \langle (\phi^{\dagger} T^{\alpha})_I \rangle \langle K^I_{I'} \rangle \langle D^{\alpha} \rangle 
 = (m^2)_{I'}^J \langle \phi^{\dagger}_J \rangle + \langle U_{I'} \rangle
\label{<VI'soft>}
\end{eqnarray}
and 
\begin{eqnarray}
&~& \langle F^I \rangle \left\langle (K_{I'}(T^{\alpha'}\phi)^{I'})_I^J \right\rangle \langle F_J \rangle 
+ \frac{1}{2} \langle (\mbox{Re} f_{\alpha \beta})_{I'} \rangle \langle (T^{\alpha'} \phi)^{I'} \rangle \langle D^{\alpha} \rangle \langle D^{\beta} \rangle  
\nonumber \\
&~& ~~~~~~~~~ + (\hat{M}_V^2)^{\alpha\alpha'} \langle D^{\alpha} \rangle 
= (m^2)_{I'}^J \langle \phi^{\dagger}_J \rangle \langle (T^{\alpha'}\phi)^{I'} \rangle~,
\label{<VI'Dsoft>}
\end{eqnarray}
respectively.
The formula (\ref{<VI'Dsoft>}) is a counterpart of (3.54) in Ref.\citen{KMY2}.

If the soft SUSY breaking terms are related to the SUSY extension of SM directly,
the magnitude of $(m^2)_{I'}^J$ should be the same size as or less than $O(1)$TeV$^2$.
In this case with $(\hat{M}_V^2)^a \gg (m^2)_{I'}^J$, the soft SUSY breaking terms are treated as a perturbation.
Then the same argument as that in the previous section is applied, 
and the same result (1) is obtained 
if $\langle F^I \rangle \left\langle (K_{I'}(T^{a}\phi)^{I'})_I^J \right\rangle \langle F_J \rangle$
is bigger than $(m^2)_{I'}^J \langle \phi^{\dagger}_J \rangle \langle (T^{a}\phi)^{I'} \rangle$.

\subsection{Case with local SUSY}

In the Einstein supergravity, the scalar potential is given by\cite{CJSFGN,CFGP}
\begin{eqnarray}
V_{\rm SG} = M^{2}e^{G/M^{2}} (G^I (G^{-1})^J_I G_J -3M^{2})
 + \frac{1}{2} \mbox{Re} f_{\alpha \beta} D^{\alpha} D^{\beta}~,
\label{scalar-potential}
\end{eqnarray}
where $M$ is a gravitational scale defined by $M \equiv M_{\rm Pl}/\sqrt{8\pi}$ using
the Planck scale $M_{\rm Pl}$, $G(\phi^I, {\phi}^{\dagger}_J)$ is the total K\"ahler potential  
defined by 
\begin{eqnarray}
G(\phi^I, \phi^{\dagger}_J) \equiv K(\phi^I, \phi^{\dagger}_J) +M^{2}\ln {|W (\phi^I) |^2 \over M^{6}} 
\label{total-Kahler}
\end{eqnarray}
and $D$-auxiliary fields are defined by
\begin{eqnarray}
D^\alpha \equiv  -(\mbox{Re} f^{-1})_{\alpha\beta} G_I ( T^\beta \phi)^I = -(\mbox{Re} f^{-1})_{\alpha\beta} (\phi^{\dagger} T^\beta)_J G^J~.
\label{DSG}
\end{eqnarray}
The $F$-auxiliary fields are given by
\begin{eqnarray}
F_J = -Me^{G/2M^{2}} (G^{-1})^I_J G_I~.
\label{FSG}
\end{eqnarray}
The scalar potential is rewritten down by
\begin{eqnarray}
   V_{\rm SG} = F^I K^J_I F_J - 3M^{4} e^{G/M^{2}}  + \frac{1}{2} \mbox{Re} f_{\alpha \beta} D^{\alpha} D^{\beta}~,
\label{scalar-potential 2}
\end{eqnarray}
where $D^\alpha$ and $F^I$ are given by (\ref{DSG}) and (\ref{FSG}), respectively.

The derivative of $V$ by $\phi^{I'}$ is given by
\begin{eqnarray}
&~& \frac{\partial V_{\rm SG}}{\partial \phi^{I'}} = G_{I'} \left( {V_F \over M^2}
   + M^{2}e^{G/M^{2}} \right) - F^I K_{II'}^{J} F_J - M e^{G/2M^{2}} G_{II'} F^I 
\nonumber \\
&~& ~~~~~~~~~~~ - \frac{1}{2} (\mbox{Re} f_{\alpha \beta})_{I'} D^\alpha D^\beta
 - (\phi^{\dagger} T^\alpha )_I G^I_{I'} D^\alpha~,
\label{VISG}
\end{eqnarray}
where $V_F \equiv F^I K^J_I F_J - 3M^{4} e^{G/M^{2}}$.
Taking its VEV and using the stationary condition, we derive the formula:
\begin{eqnarray}
&~& \langle F^I \rangle \langle K_{II'}^{J} \rangle 
   \langle F_J \rangle + m_{3/2} \langle G_{II'} \rangle \langle F^I \rangle
+ \frac{1}{2} \langle (\mbox{Re} f_{\alpha \beta})_{I'} \rangle 
   \langle D^\alpha \rangle \langle D^\beta \rangle 
\nonumber \\
&~& ~~~~~~~  +  \langle (\phi^{\dagger} T^\alpha )_I \rangle
   \langle G^I_{I'} \rangle \langle D^\alpha \rangle 
 = \langle G_{I'} \rangle 
  \left( {\langle V_F \rangle \over M^2} + m_{3/2}^2 \right)~,
\label{<VISG>}
\end{eqnarray}
where $m_{3/2}$ is the gravitino mass given by
\begin{equation}
m_{3/2}= \langle Me^{G/2M^{2}} \rangle 
= |\langle e^{K/2M^{2}} {W}/{M^2} \rangle|~.
\label{gravitino}
\end{equation}
 
By multiplying $(T^{\alpha'} \phi)^{I'}$ to (\ref{VISG}) and
using the identities derived from the gauge
invariance of the total K\"ahler potential,
\begin{eqnarray}
&~& G_{II'}(T^{\alpha'} \phi)^{I'} + G_{I'} (T^{\alpha'} )_I^{I'}
   - K^J_I(\phi^{\dagger} T^{\alpha'})_J = 0~,
\label{G-inv1}\\
&~& K_{II'}^J (T^{\alpha'} \phi)^{I'} + K^J_{I'} (T^{\alpha'})^{I'}_I
       -((\phi^{\dagger} T^{\alpha'} )_{J'}G^{J'})_I^J = 0~,
\label{G-inv2}
\end{eqnarray}
we obtain
\begin{eqnarray}
&~& {\partial V \over \partial \phi^{I'}} (T^{\alpha'} \phi)^{I'} =
      \left( {V_F \over M^2} + 2 M^2 e^{G/M^2} \right) G_{I'}(T^{\alpha'}\phi)^{I'}
       - F^I (G_{I'}(T^{\alpha'}\phi)^{I'})_I^J F_J
\nonumber \\
&~& ~~~~~~~ - \frac{1}{2} (\mbox{Re} f_{\alpha \beta})_{I'} (T^{\alpha'} \phi)^{I'} D^\alpha D^\beta
      - (\phi^{\dagger} T^{\alpha})_I G^I_{I'} (T^{\alpha'} \phi)^{I'} D^\alpha~.
\label{VISGD}
\end{eqnarray}
Taking its VEV and using the stationary condition, 
we derive the formula:\cite{JKY}
\begin{eqnarray}
&~& \langle F^I \rangle \left\langle (G_{I'}(T^{\alpha'}\phi)^{I'})_I^J \right\rangle \langle F_J \rangle 
+ \frac{1}{2} \langle (\mbox{Re} f_{\alpha \beta})_{I'} \rangle 
   \langle (T^{\alpha'} \phi)^{I'} \rangle 
   \langle D^\alpha \rangle \langle D^\beta \rangle 
\nonumber \\
&~& ~~~ + \left( (\hat{M}_V^2)^{\alpha\alpha'}
+ \left({\langle V_F \rangle \over M^2} + 2 m_{3/2}^2\right)
 \langle \mbox{Re} f_{\alpha\alpha'} \rangle \right) \langle D^\alpha \rangle = 0~.
\label{<VISGD>}
\end{eqnarray}

The VEV of $V_{\rm SG}$ is given by
\begin{eqnarray}
\langle V_{\rm SG} \rangle \equiv \langle F^I \rangle \langle K_I^J \rangle \langle F_J \rangle - 3m^2_{3/2} M^{2}
+ \frac{1}{2} \langle \mbox{Re} f_{\alpha \beta} \rangle  \langle D^{\alpha} \rangle \langle D^{\beta} \rangle~.
\label{<VSG>}
\end{eqnarray}
By the requirement of $\langle V_{\rm SG} \rangle = 0$, 
the relations $\langle F^I \rangle = O(m_{3/2}M)$ and/or $\langle D^{\alpha} \rangle =O(m_{3/2}M)$ are derived
for some components.
If the soft SUSY breaking terms are related to the SUSY extension of SM directly,
the magnitude of $m_{3/2}$ should be the same size as or less than $O(1)$TeV.
In this case with $(\hat{M}_V^2)^{a} \gg m_{3/2}^2$, the soft SUSY breaking terms are treated as a perturbation and
the same result (1) is obtained with the following upper bound 
for the magnitude of dominant SUSY breaking $F$ component,
\begin{eqnarray}
\langle D^{a} \rangle \le O(\langle F^I \rangle) \le O(m_{3/2} M)~.
\label{<D><<F>SG}
\end{eqnarray}
If the gauge symmetry breaking scale is $O(M)$, the following strong constraint is derived,\cite{JKY}
\begin{eqnarray}
\langle D^{a} \rangle \le O(m_{3/2}^2)~.
\label{<D><m2}
\end{eqnarray}
In this case, the relation $m^2_{3/2} = \frac{\langle F^I \rangle \langle K_I^J \rangle \langle F_J \rangle}{3M^2}$ holds.

By multiplying $(K^{-1})^{I'}_{I''} G^{I''}$ to (\ref{VISG}), taking its VEV and using the stationary condition,
we derive the formula:
\begin{eqnarray}
&~& \langle F^I \rangle \langle K_{II'}^J \rangle \langle F_J \rangle \langle (K^{-1})^{I'}_{I''} \rangle \langle G^{I''} \rangle 
- \langle G_{II'} \rangle \langle F^I \rangle \langle F^{I'} \rangle
\nonumber \\
&~& ~~~~~~~~~~ + \frac{1}{2m_{3/2}} \langle (\mbox{Re} f_{\alpha \beta})_{I'} \rangle 
\langle F^{I'} \rangle \langle D^{\alpha} \rangle \langle D^{\beta} \rangle 
\nonumber \\
&~& = \frac{\langle F^I \rangle \langle K_{I}^J \rangle \langle F_J \rangle}{m_{3/2}^2} \left({\langle V_F \rangle \over M^2} + m_{3/2}^2\right)
+ \langle \mbox{Re} f_{\alpha \beta} \rangle \langle D^{\alpha} \rangle \langle D^{\beta} \rangle~.
\label{<VIKSG>}
\end{eqnarray}
If the VEVs of all $F$-components vanish and $m_{3/2} \ne 0$, i.e., $\langle W \rangle \ne 0$, we obtain the relation
\begin{eqnarray}
\langle \mbox{Re} f_{\alpha \beta} \rangle \langle D^{\alpha} \rangle \langle D^{\beta} \rangle = 0~,
\label{<VIKSG>F=0}
\end{eqnarray}
which means the $D$-flat conditons, $\langle D^{\alpha} \rangle = 0$.
There exists a SUSY AdS vacuum if $\langle F^I \rangle = 0$ and $\langle W \rangle \ne 0$.
This fact is directly understood from the definition (\ref{DSG}) as follows.
{}From (\ref{FSG}), the conditions $\langle F^I \rangle = \langle F_J \rangle = 0$ for all species 
are equivalent to $\langle G^I \rangle = \langle G_J \rangle = 0$ for $\langle W \rangle \ne 0$.
Then the $D$-flat conditions are derived from (\ref{DSG}).

Finally we comment on models with the Kachru-Kallosh-Linde-Trivedi (KKLT) moduli stabilization.
In the KKLT compactification, the extra potential $V_{\rm lift}$ is introduced in order to uplift
SUSY AdS vacua to dS vacua.\cite{KKLT}
In this case, (\ref{<VISGD>}) is modified as
\begin{eqnarray}
&~& \langle F^I \rangle \left\langle (G_{I'}(T^{\alpha'}\phi)^{I'})_I^J \right\rangle \langle F_J \rangle 
+ \frac{1}{2} \langle (\mbox{Re} f_{\alpha \beta})_{I'} \rangle 
   \langle (T^{\alpha'} \phi)^{I'} \rangle 
   \langle D^\alpha \rangle \langle D^\beta \rangle 
\nonumber \\
&~& ~~~~~~~ + \left( (\hat{M}_V^2)^{\alpha\alpha'}
+ \left({\langle V_F \rangle \over M^2} + 2 m_{3/2}^2\right)
 \langle \mbox{Re} f_{\alpha\alpha'} \rangle \right) \langle D^\alpha \rangle 
\nonumber \\
&~& ~~~~~~~ = \langle \partial V_{\rm lift}/\partial \phi^{I'} \rangle \langle (T^{\alpha'}\phi)^{I'} \rangle~.
\label{<VISGD>lift}
\end{eqnarray}
The formula (\ref{<VISGD>lift}) is a counterpart of (3.7) in Ref.\citen{CJ}.
If the magnitude of new term $\langle \partial V_{\rm lift}/\partial \phi^{I'} \rangle \langle (T^{\alpha'}\phi)^{I'} \rangle$
is negligibly small compared with other terms,
the same result (1) holds.

\section{Conclusions and discussion}

We have studied the magnitude of $D$-components in a generic framework of SUSY field theory.
We have found that there exists $F$-component whose VEV is comparable to
or bigger than that of $D$-component in the absence of FI term, 
the large hierarchy in the charge spectrum and strongly interacting higher-dimensional
couplings in the K\"ahler potential, % and the gauge kinetic function,
if contributions from other terms than $F$ and $D$-terms, such as soft SUSY breaking terms
or the uplifting potential, are negligible.
If all $F$-components vanish, the SUSY is unbroken in most cases.
Hence $F$-components have the initiative in the breakdown of SUSY.

We have shown that the features of magnitude on $\langle D^{\alpha} \rangle$ and $\langle F^I \rangle$,
which are obtained through explicit models, also hold in models with
a generic K\"ahler potential and a generic gauge kinetic function
if K\"ahler potential contains no strongly interacting couplings 
and contributions from other terms than $F$ and $D$-terms are negligibly small.
Though we do not obtain completely new constraints, 
it would be meaningful to report our results and clarify our statement
because it is applicable to a broad class of SUSY field theory
including effective theories derived from a fundamental theory.

\section*{Acknowledgements}
This work was supported in part by scientific grants from the Ministry of Education, Culture,
Sports, Science and Technology under Grant Nos.~18540259 and 21244036.

%\appendix
%\section{First Appendix} %Empty argument \section{} yields `Appendix'. 
%
%\section{Second Appendix}

\end{document}